\documentclass{article}

\usepackage{microtype}
\usepackage{graphicx}
\usepackage{subcaption}
\usepackage{booktabs} 

\usepackage{hyperref}

\usepackage[accepted]{icml2026}

\usepackage{amsmath}
\usepackage{amssymb}
\usepackage{mathtools}
\usepackage{amsthm}

\usepackage{listings}
\usepackage{xcolor}

\lstdefinestyle{icmlcode}{
  basicstyle=\ttfamily\small,
  numbers=left,
  numberstyle=\tiny,
  frame=single,
  breaklines=true,
  captionpos=b
}

\usepackage[capitalize,noabbrev]{cleveref}

\theoremstyle{plain}

\theoremstyle{definition}

\theoremstyle{remark}

\usepackage[textsize=tiny]{todonotes}

\icmltitlerunning{One Simple Trick for Improving the Performance of Energy-Limited Local Inference and Training}

\begin{document}

\twocolumn[
  \icmltitle{One Simple Trick for Improving the Performance \\ of Energy-Limited Local Inference and Training}



  \icmlsetsymbol{equal}{*}

  \begin{icmlauthorlist}
    \icmlauthor{Erik Schultheis}{equal,yyy}
    \icmlauthor{Maximilian Kleinegger}{equal,yyy,xxx}
    \icmlauthor{Dan Alistarh}{yyy,comp}
  \end{icmlauthorlist}

  \icmlaffiliation{yyy}{ISTA}
  \icmlaffiliation{comp}{RedHat AI}
  \icmlaffiliation{xxx}{TU Wien}

  \icmlcorrespondingauthor{Erik Schultheis}{erik.schultheis@ist.ac.at}

  \icmlkeywords{Machine Learning, ICML, consumer GPUs, Efficiency}
  \vskip 0.3in
]




\printAffiliationsAndNotice{\icmlEqualContribution}

\begin{abstract} 
  Energy supply and heat dissipation are two of the main challenges with modern GPU deployments. While typically discussed in the context of new datacenter constructions, the same constraints also apply to small form-factor consumer devices, such as the DGX spark.
In workloads characterized by alternating compute-intensive tasks such as matmuls with memory-bound operations such as norms or cross-entropy, the compute-intensive parts might hit power and/or thermal limits and start throttling.
In this short paper, we show that \emph{chunking} the workload into smaller parts that alternate compute and memory in higher frequencies, these power and temperature spikes can be smoothed out, preventing throttling and resulting in considerably faster wall-clock time \emph{and} reduced total energy consumption. 
We present several scenarios in which this effect can be exploited on a DGX Spark with up to 20\% performance and energy improvements, and demonstrate that the same phenomenon also happens on less constrained systems, such as a multi-GPU server, albeit at significantly reduced effect size of 1-2\%.

\end{abstract}

\vspace{-1em}
\section{Introduction}

Improving the efficiency of AI systems has been a major focus of research over the past few years. Breakthroughs such as FlashAttention~\cite{dao2022flashattention,dao2023flashattention2}, LoRA/QLoRA~\cite{hu2021lora,dettmers2023qlora}, and faster decoding techniques (e.g., bitsandbytes~\cite{dettmers2022bitsandbytes}, llama.cpp, or Marlin~\cite{frantar2024marlin}) have enabled the widespread deployment of powerful models across diverse hardware platforms. However, while runtime efficiency has received significant attention, improving power and energy consumption remains a much less well-studied area. It is commonly believed that efficiency improvements naturally lead to energy savings, as energy consumption closely tracks runtime. In this paper, we present initial work demonstrating that focusing specifically on power usage patterns can provide non-trivial improvements in both power consumption and runtime efficiency.

Our work stems from a simple observation encountered while training large language models on a DGX Spark. Both during vLLM inference~\citep{vllm} and  local training using LLM-Q~\cite{schultheis2026llmq}, we noticed periodic throttling behavior manifesting as audible fan speed variations and performance fluctuations. The GPU would exhibit a rhythmic pattern: periods of high-intensity computation followed by a sudden throttling, creating a characteristic sine-wave-like power consumption profile. We tracked the pattern to be associated to the execution of the LM-head layer in large-vocabulary language models, where the workload alternates between compute-intensive matrix multiplications and memory-bound operations like cross-entropy loss computation.

Further investigation revealed an interesting phenomenon: these power spikes appear to occur when the GPU hits its power limit during the dense computation phases, triggering NVIDIA's dynamic voltage and frequency scaling (DVFS) mechanisms to reduce clock speeds. The LM-head layer, with its vocabulary dimension often exceeding 100K, creates particularly intense bursts of computation-approximately $2 \cdot \text{batch} \cdot \text{seq\_len} \cdot \text{dim} \cdot \text{vocab}$ operations in the forward pass and $4 \cdot \text{batch} \cdot \text{seq\_len} \cdot \text{dim} \cdot \text{vocab}$ in backward. When $\text{vocab} \gg \text{dim}$, as is common in modern LLMs (e.g., 152k vocabulary vs. 896 hidden dimension), these operations dominate the computational profile.

Further, we discovered that \emph{chunking}~\citep{huang2019gpipe}, a technique traditionally employed to reduce memory consumption, can effectively smooth these power spikes by interleaving compute and memory operations at finer granularity. Instead of processing entire batches through the LM-head at once, chunking splits the workload into smaller segments processed sequentially. This simple modification reduced temperature spikes by 10°C, improved performance by up to 20\% on the DGX Spark, and decreased total energy consumption by 14\%. Even on less constrained datacenter GPUs like the L40S, we observed measurable improvements of 1-2\% via chunking.

\paragraph{Contributions.}
We make three main contributions:
\begin{itemize}
    \item We identify and characterize a power-throttling phenomenon that significantly impacts performance on edge devices during LLM inference and training, providing  profiling and analysis of its root causes.
    \item We demonstrate that \textit{chunking}, traditionally a memory optimization technique, serves as an effective power-smoothing strategy that can improve both runtime and energy efficiency by up to 20\% on power-constrained devices.
    \item We provide preliminary experiments across multiple systems (DGX Spark, L40S) and frameworks (LLMQ, vLLM), along with a minimal reproduction script to facilitate further research on this topic.
\end{itemize}

\section{Related Work}

\paragraph{Power Management in GPU Computing.}
Power consumption and thermal management in GPUs have been studied extensively in the context of datacenter deployments~\cite{mei2013measurement,wang2022dynamic}. NVIDIA's power management includes the DVFS mechanism that dynamically adjusts clock frequencies based on power and thermal constraints~\cite{nvidia2024tensorrt}. Recent work has shown that nvidia-smi power measurements can be imprecise~\cite{imprecise2025}, highlighting the challenges in accurately characterizing power consumption patterns. \citet{wang2022dynamic} propose dynamic GPU energy optimization for machine-learning training workloads through online configuration search. While most existing work focuses on datacenter-scale optimizations, here we focus mainly on edge device constraints.

\paragraph{Large-Vocabulary Bottlenecks in LLMs.}
The challenge of large vocabulary sizes in language models is well-known, and several mitigations have been investigated. The LM-head projection and cross-entropy computation can dominate memory footprint and computation time, especially when vocabulary size greatly exceeds hidden dimension~\cite{schultheis2026llmq}. Several optimization techniques have been proposed: Liger's fused linear cross-entropy~\cite{liger2024} addresses this bottleneck through kernel fusion, Cut Cross-Entropy~\cite{cutce2024} reformulates the loss computation to avoid materializing the full logit matrix, and torchtune exposes \texttt{CEWithChunkedOutputLoss} for memory-efficient training. While these works recognize the LM-head as a bottleneck, they primarily focus on memory efficiency rather than power consumption patterns.

\paragraph{Chunking and Micro-batching Techniques.}
Chunking is a standard memory optimization technique. Gradient accumulation and micro-batching enable training with larger effective batch sizes than memory permits~\cite{chen2016training}. In the context of transformers, chunked attention mechanisms reduce memory complexity~\cite{child2019generating}. For inference, chunked prefill in vLLM~\cite{vllm} and similar techniques split long sequences to improve latency. LLMQ specifically implements LM-head chunking through its \texttt{--lmhead-chunks} parameter, splitting token embeddings to reduce peak memory usage~\cite{schultheis2026llmq}. However, to our knowledge, no prior work has  studied chunking as a power-smoothing technique or characterized its impact on thermal throttling behavior.

\paragraph{Energy Efficiency in Deep Learning.}
While runtime optimization has been the primary focus of efficiency research, energy consumption is gaining attention due to sustainability concerns and deployment constraints. \citet{strubell2019energy} highlighted the environmental impact of training large models. Recent work on quantization~\cite{dettmers2023qlora,frantar2022gptq}, pruning~\cite{frantar2023sparsegpt}, and efficient architectures~\cite{dao2022flashattention} indirectly reduces energy consumption by decreasing computation. However, these approaches do not address the power spike patterns we identify, which can cause throttling even when average power consumption is within limits.

Our work bridges these areas by demonstrating that power consumption patterns, not just total energy usage, significantly impact performance on edge devices. We show that chunking, beyond its memory benefits, serves as an effective technique for smoothing power consumption and preventing thermal throttling.

\section{Experiments}
\label{sec:validation}

\subsection{Training of small LLMs}
One scenario in which we have large alternating compute and memory workloads is the LM-head for large-vocabulary LLM training:  logits need to be calculated by a \texttt{dim} $\times$ \texttt{vocab} matmul, followed by memory-bound cross entropy and gradients (fused kernel); backward then follows with two more large matmuls. For small LLMs, this can make up a significant fraction of the total time.

This operation parallelizes easily over different tokens, so we can apply the chunking approach discussed above: Instead of processing all tokens at once, the current minibatch is split and processed normally until after the last transformer block. Then, it is split into small microbatches that are processed sequentially. Their results need to be combined for backward, which for activation gradients means writing to different parts of the tensor for each microbatch, and for weight gradients means accumulating to the same buffer for every microbatch. The latter means that there is genuinely more memory traffic required, so there is some trade-off beyond just more kernel launches.

In this setup, we use LLMQ~\citep{schultheis2026llmq} to run training for a Qwen2-0.5B model (i.e., 896 hidden size but 152k vocabulary) and measure the total energy consumption at the wall socket. The 128GB memory of the spark supports a minibatch size of 64 sequences of 1024 tokens each. Running for 20 minutes 45 seconds consumed a total of 57 Wh, corresponding to an average power draw of 165W.
In contrast, when handling the LM-head with a microbatch of 16 instead of 64, the running time decreases to only 16 minutes 30 seconds at 49 Wh, corresponding to an increased power draw of 178W.

\subsection{Minimal reproduction}
We can also produce a minimal script intended to induce the effect:
This runs a matrix multiplication, softmax as a memory-bound operation, and another matrix multiplication.

\begin{figure}[!t]
    \centering
\begin{lstlisting}[language=Python, style=icmlcode,caption={Minimal reproduction of the chunked workload.}, label={code:minimal_reproduction}, captionpos=b]
@torch.compile
def memory_bound(c):
    c[...] = torch.softmax(c, dim=-1)

def workload_chunked(a, b, c, d, chunks: int):
    cs = M // chunks
    for m in range(0, M, cs):
        slice = torch.matmul(a[m:m+cs, :], b.T, out=c[m:m+cs, :])
        memory_bound(slice)
        d = torch.addmm(d, slice.T, a[m:m+cs, :], out=d)
    return c, d
\end{lstlisting}
\end{figure}

Since we wanted to isolate the problem and further profile and evaluate it, we implemented a simple script that reproduces the same behavior we observed during training. As shown in Listing~\ref{code:minimal_reproduction}, we simulate the LM-head training workload by alternating large matrix multiplications with a memory-bound operation. This follows the same pattern as the LM-head in large-vocabulary LLM training. In our minimal reproduction, we model the memory-bound part using a softmax over a reasonably sized matrix. To evaluate the effect of chunking, we implement a basic chunked version using a simple for loop.

Generally, we run the script for multiple numbers of chunks $\{1, 2, 4, 8, 16, 32\}$ for the multiplication of $A^{M\times K} \times B^{K\times N}$ with $M=32768$, $K=2048$, and $N=131072$. We run our experiments on a DGX Spark, considered local hardware, and on L40 GPUs to evaluate whether the effect is noticeable. We use 10 warmup steps and 100 iterations of the multiplication.

\begin{figure}[!t]
    \centering
    \includegraphics[width=\linewidth]{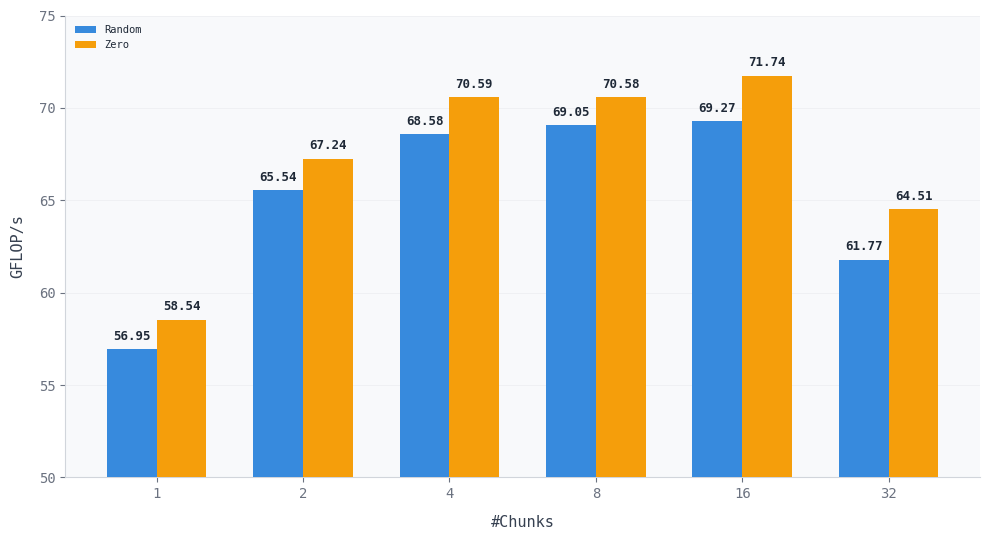}
    \caption{FLOPS per chunking configuration}
    \label{fig:chunked_perf}
\end{figure}

\paragraph{Results on DGX Spark.} This bottleneck really becomes a problem on devices with slow memory bandwidth. We can choose the DGX Spark as an example. We start by comparing the FLOPS for each number of chunks to see any improvements for this operation. As we can see in Figure~\ref{fig:chunked_perf}, using only two chunks improves our FLOPs by 15\%, reaching up to 21.63\% for chunk size 16, before the overhead of using too many chunks slows it down again. We further managed to isolate the problem by using zero-valued matrices, as seen in Figure~\ref{fig:chunked_perf} with the orange bars. We even further improve by another 1\% to 22.54\% compared to the baseline.

\begin{figure*}[!t]
    \centering
    \includegraphics[width=\textwidth]{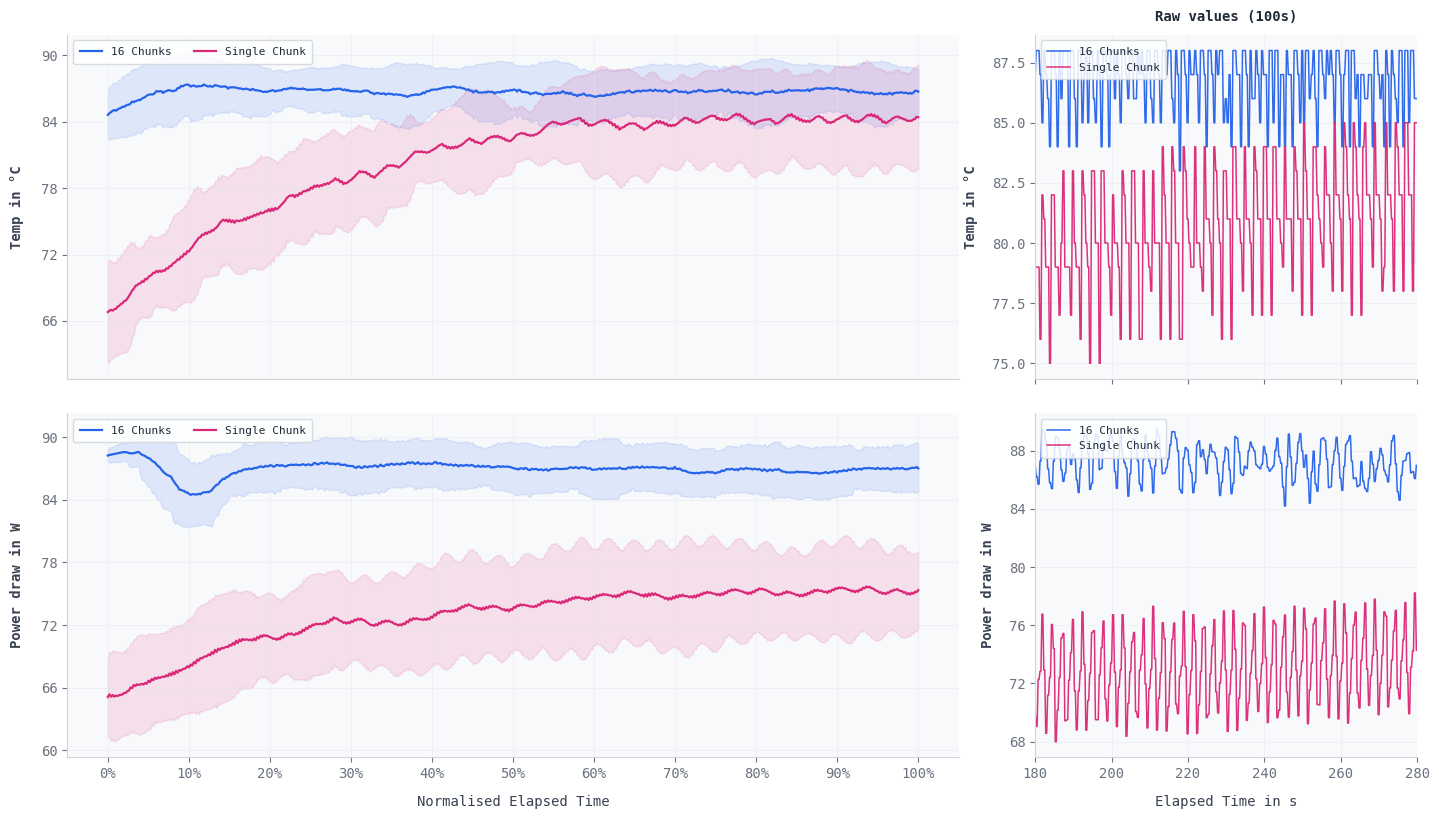}
    \caption{Temperature and power draw comparison between single-chunk and 16-chunk execution. The plots on the left show normalized telemetry over the full iteration together with the corresponding confidence intervals. The plots on the right show the raw telemetry values for a representative 100-second sequence.}
    \label{fig:temp_powerdraw}
\end{figure*}

Furthermore, we want to consider temperature and power draw, as those should correlate and potentially be the reason for the performance throttling. In Figure~\ref{fig:temp_powerdraw}, we plot the single-chunk configuration and the best-performing one (16 chunks). As we can see, the overall temperature for the single-chunk sequence is lower by around 10°C, but the confidence interval is almost twice as high compared to 16 chunks. Considering the right side, we see that power draw and therefore temperature are more stable when using chunking compared to using a single chunk. Therefore, those unstable spikes in power draw overall slow down the execution.

\begin{figure}[!t]
    \centering

    \includegraphics[width=\linewidth]{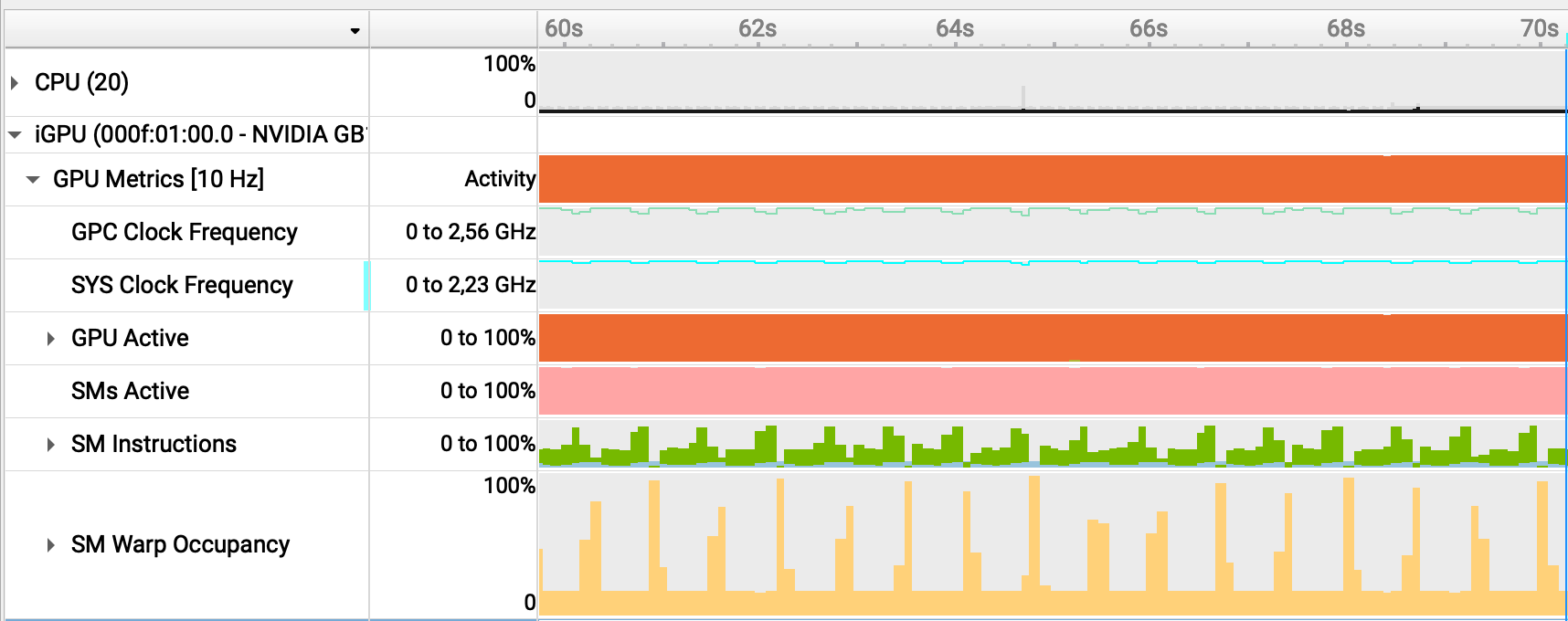}

    \vspace{0.5em}
    \includegraphics[width=\linewidth]{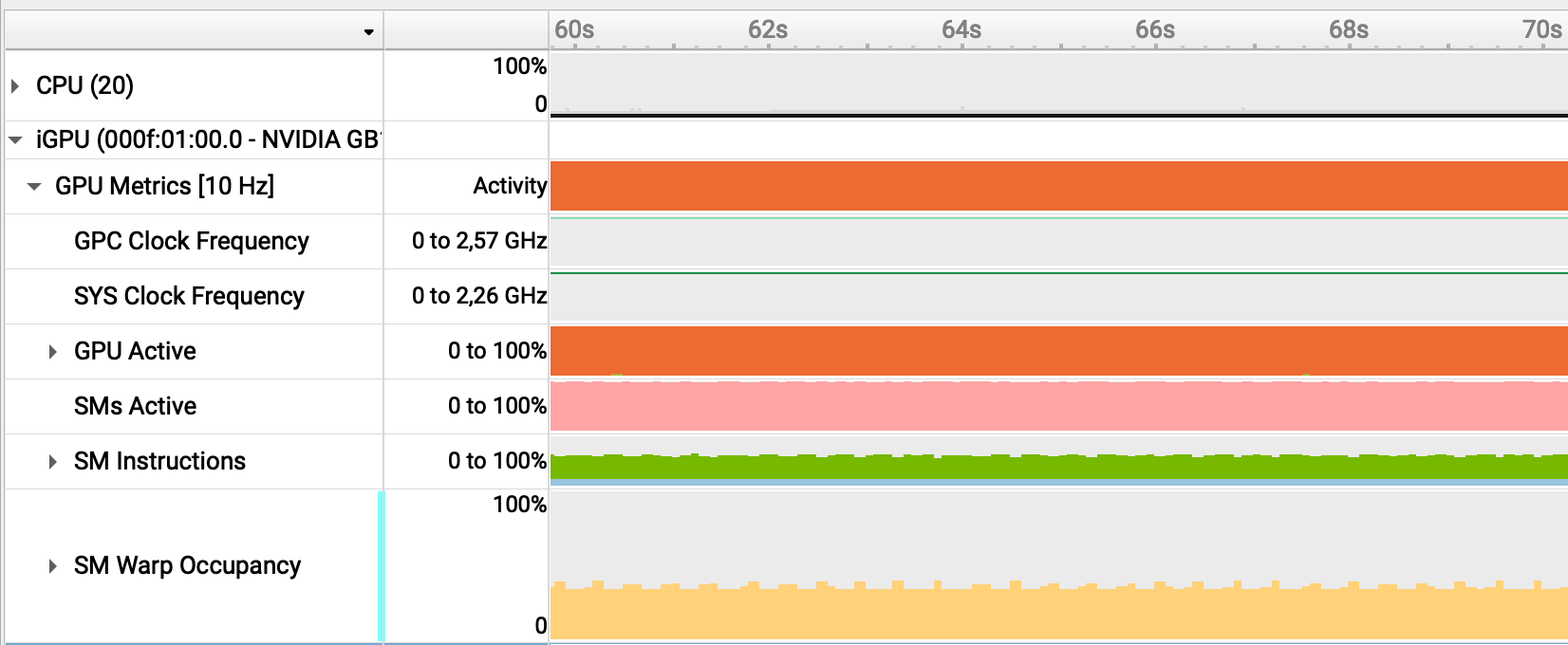}

    \caption{Nsight profiling comparison between single-chunk (above) and 16-chunk (below) execution.}
    \label{fig:nsight_comparison}
\end{figure}

Lastly, we want to dive deeper to evaluate where those spikes originate. For this, we profiled our script using Nsight Systems~\cite{nvidia_nsight}. We collected GPU metrics and present them with equal filtering and zooming in Figure \ref{fig:nsight_comparison}. Instantly, we notice that the single-chunk setup has a more periodic warp occupancy compared to the 16-chunk version. This potentially also leads to some GPC and SYS clock variations in the single-chunk setup, compared to stable clocks in the 16-chunk setting. This leads us to conclude that using chunks, the warp scheduler manages to keep warps busy continuously with a more uniform distribution, avoiding the periodic bursts of intense warp execution seen in the single-chunk case, which should lead to higher power draw and therefore higher temperature.

\paragraph{Results on L40S GPUs.}
For an L40S, the same effect can be observed, but at strongly diminished strength:
Running in a single chunk resulted in a duration per iteration of 796 ms, whereas using four chunks, the best observed configuration, decreased this number to 776 ms. At the same time, energy consumption decreased from 2747 J to 2712 J and power draw increased from 344.8 W to 349.5 W.
Thus, we achieve a 2.5\% improvement in timing and 1.3\% in energy.

If we change all operands to be zero matrices, the effect almost goes away. Timing changes from
631 ms to 626 ms, and energy consumption from 1687 J to 1682 J corresponding to 267 W and 269 W, respectively. By using all-zero matrices, dynamic power draw required for flipping transistor states reduces, moving the operating point into a less power-constrained regime. Consequently, the speedup due to additional chunking-based smoothing reduces to only 0.7\%.

\subsection{Performance Impact during Inference}

Naturally, after evaluating the performance during sequences of operations that mostly occur during LLM training, we asked ourselves whether a similar effect could also happen during inference. For this reason, we evaluated a similar experimental setup as before using vLLM~\cite{vllm} to measure prefill performance. We used their latency benchmark with 3 warmup steps and 5 iterations to gather latency measurements. We evaluated this on an 8B Llama model~\cite{dubey2024llama}.

\begin{figure}
    \centering
    \includegraphics[width=1\linewidth]{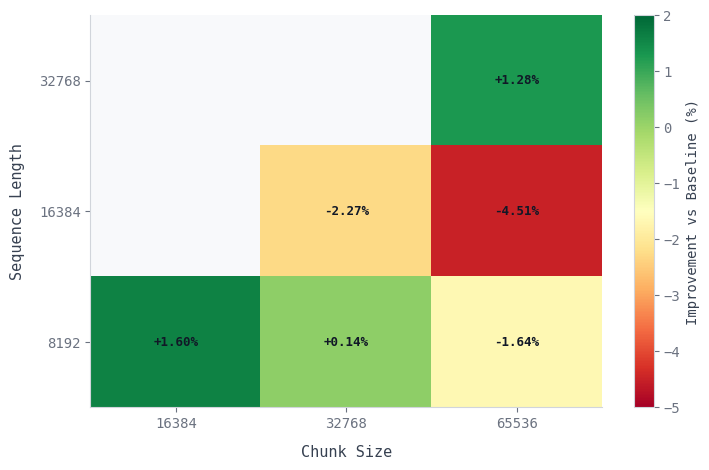}
    \caption{Improvements over the baseline under different sequence lengths and chunk sizes in vLLM.}
    \label{fig:inference}
\end{figure}

At first, we did not manage to see any effect at all, but this was due to using too small sequence lengths. After increasing them towards 32k and using proper chunk sizes, we started to see improvements. However, those improvements were relatively small, reaching at most 1.6\%. More importantly, the overhead of chunking the prefill can easily slow down the execution and therefore decrease performance. Therefore, this needs to be considered and evaluated carefully. The results can be seen in Figure~\ref{fig:inference}.

Profiling vLLM, on the other hand, is rather difficult, since it is already highly optimized and a lot is happening internally. Therefore, we did not manage to isolate and evaluate the effect properly.

\section{Discussion}
\label{sec:discussion}

\paragraph{Why does chunking help?}
Our measurements indicate that the benefit of chunking on power-constrained devices stems from smoothing the \emph{instantaneous} power draw, not from any change in total arithmetic.
When a large GEMM is issued as a single kernel, the device's SMs saturate near-simultaneously and produce a sharp power transient; if this transient pushes against the power or thermal envelope, the firmware reacts via DVFS, lowering core clocks until conditions return to a safe regime.
Once the kernel completes and a memory-bound kernel runs, power drops abruptly and the device cools, but the boost cycle has already cost performance.
Chunking interleaves a small matmul with a memory-bound operation many times in succession, producing a far flatter power profile (Fig.~\ref{fig:temp_powerdraw}, right).
The Nsight traces in Fig.~\ref{fig:nsight_comparison} corroborate this picture: warp occupancy and clock frequencies are visibly more uniform under chunking.
As a result, the device sustains its boost clocks longer and avoids reactive frequency dips, which more than offsets the modest overhead introduced by additional kernel launches and gradient accumulation.

\paragraph{Why edge devices benefit more.}
The improvement we measure on the DGX Spark (15-22\% on the kernel benchmark; up to 20\% wall-clock and 14\% energy on end-to-end training) is markedly larger than on the L40S (1-2\%).
This is consistent with the fact that small form-factor systems have substantially tighter power and thermal envelopes than rack-mounted accelerators: the Spark dissipates its budget into a quiet, compact enclosure, whereas the L40S sits in a server with aggressive cooling and ample TDP headroom.
Because firmware engages DVFS only when those envelopes are pressed, the same workload triggers throttling far more frequently on the Spark.
We expect the trend to extend to other consumer-facing platforms (laptops, Jetson-class devices, and workstation-grade RTX cards), where the cost of crossing the envelope is high but is rarely modeled by performance estimators built for datacenter hardware.

\paragraph{Practical implications.}
Two implications stand out.
First, training and inference frameworks intended for edge or consumer deployment should expose chunking parameters not only as a memory knob but also as a tool for power-shape optimization; LLMQ already supports this~\cite{schultheis2026llmq}, but most other frameworks do not.
Second, performance models that estimate runtime from FLOPS and memory bandwidth alone are misleading on power-limited platforms: identical FLOP counts can produce very different wall-clock times depending on how the computation is paced.

\paragraph{Limitations.}
Our study is preliminary in several respects.
We focus on a single edge device (DGX Spark) and one datacenter GPU (L40S); intermediate platforms such as the RTX 5090, Jetson AGX, or Apple silicon may behave differently.
Our end-to-end training experiment is limited to a small Qwen2 variant, and the scaling behavior of the chunking benefit at larger model sizes remains an open question.
Power measurements via NVML are known to be imprecise~\cite{imprecise2025}, so we cross-check with wall-socket readings and profiler traces, but cannot fully decouple thermal from electrical throttling.
Finally, the optimal chunk size depends on the workload, the device, and the surrounding kernels; we do not yet provide an automatic selection mechanism.

\paragraph{Future work.}
A natural extension is online adaptive chunking, in which a controller monitors SM clocks or instantaneous power and selects chunk sizes that maximize sustained throughput.
Combining chunking with kernel-fused alternatives such as Liger~\cite{liger2024} or Cut Cross-Entropy~\cite{cutce2024} is also promising, since fused kernels reduce launch overhead but can themselves be partitioned into power-friendly chunks.
Finally, the same principle should generalize beyond LM-heads to any workload that alternates between arithmetic-intensive and memory-bound phases; MoE routing, large QK projections, and convolutional bottlenecks are natural candidates we leave to future investigation.

\section{Conclusion}
\label{sec:conclusion}
{We showed that, on power- and thermally-constrained GPUs, periodic throttling can lead to lost performance and wasted energy in workloads that alternate compute- and memory-bound phases.
We showed that \emph{chunking}, a standard memory optimization, can also be a remarkably effective power-smoothing technique: on a DGX Spark, applying it to the LM-head reduces training wall-clock by 20\% and total energy by 14\%, while on the less constrained L40S the effect is modest but still measurable.
We also provide a minimal Python reproduction that isolates the same phenomenon on a synthetic matmul-softmax-matmul workload and confirms that the underlying mechanism is improved power-draw stability rather than any change in arithmetic intensity, and can help other researchers reproduce and study this phenomenon.
We hope this short paper can lead to additional attention to the underexplored interaction between scheduling granularity and power management.}

\bibliography{bib}
\bibliographystyle{icml2026}

\end{document}